\documentclass[aps,pra,10pt,twocolumn,balancelastpage,showpacs]{revtex4-1}
\usepackage{times}
\usepackage{amsmath}
\usepackage{amssymb}
\usepackage{graphicx}
\graphicspath{{figures/}{/}{../resubm2-pre/}}
\usepackage{verbatim}

\newcommand{\eqn}  {Eq.~}

\newcommand  {\avg}[1]        {\langle#1\rangle}

\newcommand  {\ee}            {\mathrm{e}}
\newcommand  {\ii}            {\mathrm{i}}
\newcommand  {\nup}           {N_{\mathord{\uparrow}}}
\newcommand  {\nb}            {N_\mathrm{b}}
\newcommand  {\gammabar}      {\overline{\gamma}}

\newcommand  {\avglta}[1]     {\overline{\langle#1\rangle(t)}}

\newcommand  {\ket}[1]        {\lvert#1\rangle}

\newcommand  {\braket}[2]     {\langle#1\vert#2\rangle}
\newcommand  {\bramidket}[3]  {\langle#1\vert#2\vert#3\rangle}
\newcommand  {\abs}[1]        {\lvert#1\rvert}

\newcommand  {\var}           {\mathop{\mathrm{var}}\nolimits}
\newcommand  {\Tr}            {\mathop{\mathrm{Tr}}\nolimits}

\begin{document}

\title{Off-diagonal matrix elements of local operators in many-body quantum systems}

\author{Wouter Beugeling} 
\author{Roderich Moessner} 
\author{Masudul Haque}

\affiliation{Max-Planck-Institut f\"ur Physik komplexer Systeme, N\"othnitzer Stra\ss e 38, 01187 Dresden, Germany}

\date{\today}
%% PACS
%05.30.-d   Quantum statistical mechanics
%05.70.Ln   Thermodynamics -- Nonequilibrium and irreversible thermodynamics
%75.10.Pq   General theory and models of magnetic ordering -- Spin chain models 

%% abstract
\begin{abstract}

  In the time evolution of isolated quantum systems out of equilibrium, local observables generally
  relax to a long-time asymptotic value, governed by the expectation values (diagonal matrix
  elements) of the corresponding operator in the eigenstates of the system.  The temporal
  fluctuations around this value, response to further perturbations, and the relaxation toward this
  asymptotic value, are all determined by the off-diagonal matrix elements.  Motivated by this
  non-equilibrium role, we present generic statistical properties of off-diagonal matrix elements of
  local observables in two families of interacting many-body systems with local interactions.  Since
  integrability (or lack thereof) is an important ingredient in  the relaxation process,
  we analyze models that can be continuously tuned to integrability.  We show that, for generic
  non-integrable systems, the distribution of off-diagonal matrix elements is a gaussian centered at
  zero.  As one approaches integrability, the peak around zero becomes sharper, so that the
  distribution is approximately a combination of two gaussians.  We characterize the proximity to
  integrability through the deviation of this distribution from a gaussian shape.  We also determine
  the scaling dependence on system size of the average magnitude of off-diagonal matrix elements.

\end{abstract}

\maketitle

\section{Introduction}

The topic of non-equilibrium dynamics of thermally isolated quantum systems has enjoyed a resurgence
of interest, partly because of experimental progress with cold atoms.  An isolated system has no
relaxation mechanism toward the low-lying parts of the many-body spectrum.  As a result, the
properties of eigenstates far from the edges of the spectrum may be more important for a
non-equilibrium experiment than the low-energy parts of the spectrum, which is the traditional focus
of interest of many-body quantum theory.

A key question in the non-equilibrium dynamics of isolated quantum systems is the thermalization or
relaxation of a system prepared far out of equilibrium and subject to a time-independent
Hamiltonian.  The value (if any) to which local observables relax is determined by the diagonal
matrix elements $A_{\alpha\alpha}=\bramidket{\psi_\alpha}{\hat{A}}{\psi_\alpha}$ of the
corresponding operator $\hat{A}$ in the eigenstates $\ket{\psi_\alpha}$.  The eigenstate
thermalization hypothesis (ETH)
\cite{Deutsch1991,Srednicki1994,RigolEA2008,PolkovnikovEA2011,GemmerEA2009book} proposes that the
mechanism for the thermalization of non-integrable (``chaotic'') systems is the smoothness of
$A_{\alpha\alpha}$ as a function of eigenenergies $E_{\alpha}$.  Accordingly, diagonal matrix
elements of local operators have been the subject of several studies
\cite{RigolEA2008,Rigol2009PRL,RigolSantos2010,BiroliEA2010,IkedaEA2011,IkedaEA2013,
  RigolSrednicki2012,SteinigewegEA2013,NeuenhahnMarquardt2012,BeugelingEA2014PRE, KotaEA2011,
  Motohashi2011}.

Off-diagonal matrix elements, $A_{\alpha\beta}=\bramidket{\psi_\alpha}{\hat{A}}{\psi_\beta}$,
provide further information about the time evolution $\avg{A}(t)$ of observables.  In any finite
system initially prepared in a combination of many eigenstates, there will be residual temporal
fluctuations around the long-time average.  These temporal fluctuations have been the subject of
several recent studies, both numerical \cite{Rigol2009PRA,KhatamiEA2012,GramschRigol2012,HeEA2013,
  ZangaraEA2013,ZiraldoSantoro2013, RigolEA2008} and analytical
\cite{Reimann2008,Short2011,VenutiZanardi2013}.  The magnitude of these fluctuations is determined
by $\abs{A_{\alpha\beta}}^2$, weighted, of course, by the weights of the eigenstates in the
non-equilibrium initial state.  Autocorrelation functions (unequal-time correlators), interesting on
their own and appearing in the formulation of fluctuation-dissipation relations in the ``relaxed''
state a long time after a quench \cite{KhatamiEA2013}, also are given in terms of
$\abs{A_{\alpha\beta}}^2$.  Finally, the details of the temporal approach to the final relaxed value
are also determined by the off-diagonal matrix elements of the corresponding operator
\cite{Rigol2009PRA,BrandinoEA2012}.  The approach toward the final value has been calculated in some
model systems \cite{BanulsCiracHastings2011, GenwayEA2012, GenwayEA2013,
  FagottiColluraEsslerCalabrese_PRB2014, FagottiEssler2013PRB}, although the connection to
off-diagonal matrix elements has not been explored in detail.

The (statistical) properties of off-diagonal matrix elements of local operators, $A_{\alpha\beta}$,
are thus related to a range of temporal properties of contemporary interest.  In this work, we
provide a statistical study of these objects.  We use Hamiltonians that can be tuned between
integrable limits, and provide scaling analyses as a function of system size.  We thus study what
happens to the distributions of $A_{\alpha\beta}$ as a function of distance from integrability, as
well as how the thermodynamic limit is approached.

Some statistical aspects of off-diagonal matrix elements $A_{\alpha\beta}$ have appeared in Ref.\
\cite{KhatamiEA2013} in the context of a non-equilibrium fluctuation-dissipation relation, and in
Ref.~\cite{SteinigewegEA2013}.  The aim of the present paper is to focus directly on the
$A_{\alpha\beta}$ in a manner independent of quench protocol and provide a thorough study of their
statistical properties.

In the time evolution $\avg{A}(t)$, each matrix element $A_{\alpha\beta}$ contributes with a
frequency equal to the eigenvalue difference $E_\beta-E_\alpha$
\cite{Srednicki1999,ZiraldoEA2012,KhatamiEA2013}.  In many quenches of physical interest, the
initial occupancies are confined to a small energy window (e.g.,
\cite{RigolEA2008,SantosEA2012,TorresHerreraSantos2013,*TorresHerreraSantos2014}), yet involve many
eigenstates \cite{LindenEA2009}.  We therefore pay particular attention to the behavior of the
typical values of $A_{\alpha\beta}$ for small $E_\beta-E_\alpha$.  At large frequencies, the average
$\abs{A_{\alpha\beta}}$ falls off fast, exponentially or super-exponentially with
$E_{\beta}-E_{\alpha}$.

We pay special attention to the proximity to integrability, since it is well-appreciated that the
relaxation behavior of chaotic or generic systems is quite different from systems subject to
integrability \cite{RigolEA2007,BarthelSchollwock2008,IucciCazalilla_PRA09,
  Rigol2009PRL,SantosRigol2010,CassidyEA2011,CazalillaIucciChung_PRE12, HeRigol2012,HeEA2013,
  ColluraEA2013, FagottiColluraEsslerCalabrese_PRB2014, FagottiEssler2013JStatMech,
  Fagotti2013,CauxEssler2013,ColluraEA2014} or to (many-body) localization
\cite{SantosRigol2010,CanoviEA2011,ZiraldoEA2012,ZiraldoSantoro2013,SirkerEA2014,
  NandkishoreHuse2014preprint,HickeyEA2014preprint}.
We identify signatures of the $A_{\alpha\beta}$ typical to the integrable, close-to-integrable, and
nonintegrable cases.  Close to integrability, we show that the matrix $\abs{A_{\alpha\beta}}$ has a
block-like or banded structure as a function of the energy difference (frequency)
$E_{\beta}-E_{\alpha}$, which is visible as oscillatory behavior in the frequency-dependence of
average $\abs{A_{\alpha\beta}}^2$ values.

We show that the distribution of the matrix elements in any small frequency window is peaked around
zero, having a near-gaussian form for generic non-integrable systems (cf.\ Ref.\
\cite{SteinigewegEA2013}).  At or near integrability, there is a stronger peak around zero, i.e.,
the probability distribution is a mixture of two gaussian-like curves with unequal widths.  This
difference appears to be a basic distinction between generic (non-integrable) and integrable
systems.  We demonstrate how the proximity to integrability can be quantitatively characterized
through the shape of the distribution of $A_{\alpha\beta}$ values, e.g., through the size dependence
of the kurtosis of this distribution.  

We find that the scaling behavior of the average value of $\abs{A_{\alpha\beta}}^2$ is $D^{-1}$ in
terms of the Hilbert-space dimension $D$.  The values of $\abs{A_{\alpha\beta}}^2$ at low
frequencies tend to be larger than for the generic matrix elements, but the scaling follows $D^{-1}$
as well.  The scaling analysis is analogous to studies of the diagonal matrix elements
$A_{\alpha\alpha}$ and related quantities as a function of system size, performed, e.g., in
Refs.\ \cite{DubeyEA2012,IkedaEA2013, SteinigewegEA2013,SteinigewegEA2014,
  BeugelingEA2014PRE,KhodjaEA2014preprint}.  As for the diagonal fluctuations
\cite{NeuenhahnMarquardt2012,BeugelingEA2014PRE}, we can construct plausibility arguments based on
an assumption of quasi-randomness of the vector coefficients of the energy eigenstates.  As such
assumptions are difficult to prove rigorously, we emphasize, as in Ref.\ \cite{BeugelingEA2014PRE},
that such arguments are inherently heuristic and that extensive, multi-system, numerical analysis is
required to establish scaling laws; this paper provides such data.

The size dependence of $A_{\alpha\beta}$'s is related to the size dependence of the magnitude of the
temporal fluctuations around the long-time average \cite{GramschRigol2012,HeEA2013, ZangaraEA2013}.
The $D^{-1}$ scaling is consistent with the exponential dependence of the long-time fluctuations on
the system size \cite{ZangaraEA2013}.

This paper is structured as follows. 
In Sec.~\ref{sect_models_observables}, we introduce our models: the XXZ ladder and the Bose-Hubbard
chain.  In Sec.~\ref{sect_spectral}, we introduce the frequency-resolved average of the off-diagonal
matrix elements.
In Sec.~\ref{sect_twogaussian}, we analyze the distribution of values of $A_{\alpha\beta}$,
characterizing how a mixed distribution (with two components having different widths) emerges close to
integrability.
Sec.~\ref{sect_scaling} provides a scaling analysis of the size-dependence of the average values of
$\abs{A_{\alpha\beta}}^2$, focusing on the low-frequency matrix elements.
Sections \ref{sect_spectral}, \ref{sect_twogaussian}, and \ref{sect_scaling} show results for the XXZ
ladder.  We support the generality of these results by presenting corresponding data for the
Bose-Hubbard chain in Sec.~\ref{sect_bosehubbard}.  
In the appendices, we provide details of the relationship between time evolution $\avg{A}(t)$ and
the matrix elements $A_{\alpha\beta}$, and about our quantification of the non-gaussian
distributions.

\section{Models and observables}%
\label{sect_models_observables}%

We use two families of Hamiltonians, each of which can be tuned to integrable points.  Both have been
used in our previous work on diagonal matrix elements \cite{BeugelingEA2014PRE}.  Because we are
interested in generic properties of matrix elements, we take care to avoid spurious symmetries in
our model systems.

In the spirit of many thermalization studies using spin models
\cite{SantosEA2012,MarcuzziEA2013,SteinigewegEA2013,FagottiEssler2013JStatMech,NiemeyerEA2014,
  FagottiColluraEsslerCalabrese_PRB2014,ZiraldoEA2012,SteinigewegEA2014}, our first tunable model will be the
spin-$\tfrac{1}{2}$ Heisenberg XXZ ladder with the geometry introduced in
Ref.~\cite{BeugelingEA2014PRE}. One ladder leg has an extra site compared to the other.  There are
thus $L=2p+1$ sites, with $p$ rungs between the legs.  This geometry avoids reflection symmetries.
We have nearest-neighbor Heisenberg couplings
\begin{equation}
  h_{i,j} = \tfrac{1}{2}\left(S_i^+S_j^- + S_i^-S_j^+\right) + \Delta S_i^z S_j^z,
\end{equation}
with $S_i^\pm = S_i^x\pm\ii S_i^y$, where $S_i^\mu$ ($\mu=x,y,z$) are the spin operators, and $i,j$
denote the nearest-neighbor site pairs. The anisotropy parameter $\Delta$ is kept away from special
values like $0$ and $\pm 1$, in order to avoid $SU(2)$ symmetry or special solvable points; we use
$\Delta=0.8$.  The Hamiltonian of the system is $H=H_0+\lambda H_1$, where
\begin{equation}\label{eqn_h_xxzladder}
  H_0=\sum_{i=1}^{p-1}h_{i,i+1} + \sum_{i=p}^{2p}h_{i,i+1}
  \qquad\text{and}\qquad
  H_1=\sum_{i=1}^{p}h_{i,i+p}
\end{equation}
are the intrachain (leg) and the interchain (rung) coupling, respectively. The rung coupling is
multiplied by $\lambda$, which acts as a tuning parameter.  
The $xy$ coupling along the ladder legs sets the units of energy and frequency.  
For $\lambda=0$, the chains are
uncoupled and the model is integrable. For finite values of $\lambda$, the system is
non-integrable. In the limit of large $\lambda$, where the rung couplings dominate, there is another
integrable limit. The effect of varying $\lambda$ on the fluctuations of diagonal matrix elements
has been studied in detail in Ref.\ \cite{BeugelingEA2014PRE}.

The number $\nup$ of up spins is a conserved quantity. The analysis can therefore be constrained to
a fixed-$\nup$ sector.  The dimension of the Hilbert space of the $(L,\nup)$ sector is equal to the
binomial coefficient $D=\binom{L}{\nup}$.  In order to study scaling, we use a sequence of
system sizes with almost constant filling fraction.  We present data for a sequence of systems with
near-zero magnetization (near half filling), by choosing $L=2p+1$ and $\nup=p$ for integer $p$.

Discussion of thermalization generally concerns local observables.  We present data for $S^z_2$ and
$S^z_2S^z_{p+2}$, which serve as representative examples of single-site and two-site operators.

%%%%%%%%%%%%%%% FIGURE %%%%%%%%%%%%%%%
\begin{figure*}[tb]
\noindent\center
\includegraphics[width=0.95\textwidth]{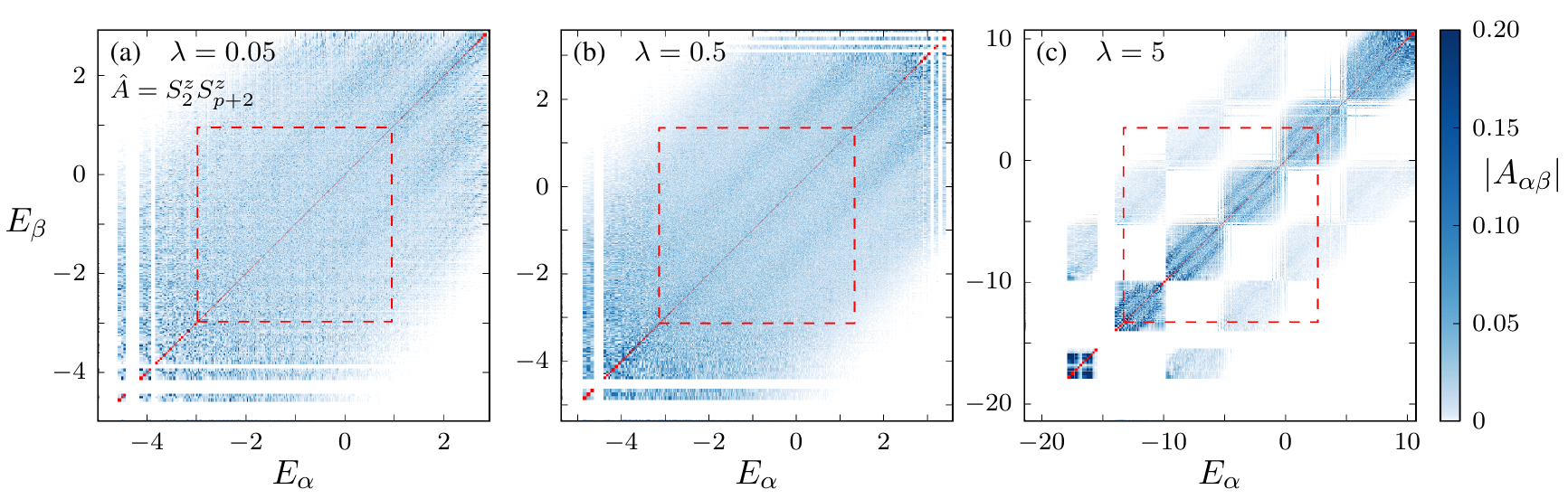}
\caption{  \label{fig_Aalphabeta}
(Color online) Matrix structure of $\abs{A_{\alpha\beta}}$ as function of $E_\alpha$ and $E_\beta$ for
the observable $\hat{A}=S^z_2S^z_{p+2}$.  The diagonal matrix elements are ignored. The white
bands near the edges are regions without eigenvalues. The dashed square indicates the
central half of the energy range, i.e.,
    $\left[\frac{3}{4}E_\mathrm{min}+\frac{1}{4}E_\mathrm{max},
      \frac{1}{4}E_\mathrm{min}+\frac{3}{4}E_\mathrm{max}\right]$ in each direction: this is the
``bulk'' of the  spectrum on which we focus our analysis.  The system size is $(L,\nup)=(13,6)$; the
Hilbert space dimension is $D=1716$.  The unit of energy is set by the $xy$ coupling along the ladder
legs.  
}
\end{figure*}
%%%%%%%%%%%%%%%%%%%%%%%%%%%%%%%%%%%%%%

The second tunable Hamiltonian is the Bose-Hubbard model, widely used in studies of thermalization
\cite{KollathLaeuchliAltman_PRL07,
  Roux2009,Roux2010PRA81,ZhangEA2011,ZhangEA2012PRA,SorgEA2014preprint, BiroliEA2010, CramerEA2008,
  CramerEisert2010}.  We use the Bose-Hubbard Hamiltonian on an $L$-site chain, with an extra term
at an edge site killing reflection symmetry, as in Ref.~\cite{BeugelingEA2014PRE}:
\begin{equation}\label{eqn_h_bosehubbard}
  H_\mathrm{BH} = \sum_{i=1}^{L-1} \left(b_i^\dagger b_{i+1} + b_{i+1}^\dagger b_{i}\right) + \lambda \left(\sum_i b_i^\dagger b_i^\dagger b_i b_i + H_\Delta \right)
\end{equation}
where $b_i$ is the creation operator at site $i$ and $H_\Delta = \Delta b_1^\dagger b_1^\dagger b_1
b_1 $ with $\Delta=0.1$ is a small perturbation to the interaction term at the first site. The
system is integrable in the $\lambda\to0$ and $\lambda\to\infty$ limits, and nonintegrable for
intermediate values.  We show results for the sector of unit filling fraction, i.e., the number of
bosons $\nb=L$. This choice provides the same sequence of Hilbert-space sizes [for bosons,
$D=\binom{L+\nb-1}{\nb}$] as the one given by $L=2\nup+1$ for the XXZ ladder. Typical local
observables in the study of this model include $n_i=b^\dagger_i b_i$, $b^\dagger_i b_{i+1} +
b^\dagger_{i+1} b_{i}$ and $n_i n_{i+1}$.

%%% FIGURE 1  TAKEN FROM HERE %%%

\section{Frequency-resolved average matrix elements}%
\label{sect_spectral}%

In Fig.~\ref{fig_Aalphabeta}, we visualize through a density plot the structure of the matrix
$\abs{A_{\alpha\beta}}$ as a function of energies $E_\alpha$ and $E_\beta$, using the rung correlator
$\hat{A}=S^z_2S^z_{p+2}$ of the XXZ ladder as observable.  The diagonal matrix elements are not
considered.  The structure of darker bands parallel to the main diagonal suggests that the magnitude
of the $\abs{A_{\alpha\beta}}$ depends roughly on the difference $E_\alpha-E_\beta$.  Thus the
energies $(E_\alpha,E_\beta)$ rather than the indices $(\alpha,\beta)$ are natural coordinates
for this plot (cf.\ Refs.~\cite{Rigol2009PRA,GenwayEA2012}). 

To consider the $\abs{A_{\alpha\beta}}$ from finite-size data as a continuous function of frequency,
we ``smooth out'' $\abs{A_{\alpha\beta}}^2 \delta\left(\omega-(E_\beta-E_\alpha)\right)$ as a
function of $\omega$, by averaging the values of $\abs{A_{\alpha\beta}}^2$ with $E_\alpha-E_\beta$
in the frequency window $[\omega-\Delta\omega,\omega+\Delta\omega]$,
\begin{equation}\label{eqn_sa2_of_omega}
  S_A^2(\omega,\Delta \omega)
  \equiv \frac{1}{\tilde{N}_{\omega,\Delta \omega}}
    \sum_{\substack{\alpha,\beta\\ \alpha\not=\beta\\ E_\alpha-E_\beta\in[\omega-\Delta \omega,\omega+\Delta \omega]}}\abs{A_{\alpha\beta}}^2,
\end{equation}
where $\tilde{N}_{\omega,\Delta \omega}$ is the number of state pairs satisfying
$E_\alpha-E_\beta\in[\omega-\Delta \omega,\omega+\Delta \omega]$. The frequency-window width
$2\Delta\omega$ is chosen such that the interval contains sufficiently many pairs of states. 
We restrict ourselves to positive $\omega$, since $A_{\alpha\beta}=A_{\beta\alpha}$ for hermitian
observables.
The quantity $S_A^2(\omega)$ is closely related to fluctuations around the asymptotic
value to which $\avg{A(t)}$ relaxes a long time after a quantum quench
(Appendix~\ref{app_time_evolution}).
The quantity $S_A(\omega,\Delta\omega)$ is the standard deviation of the distribution formed by the
$A_{\alpha\beta}$ in the frequency window.  

\begin{figure}[tb]
\centering
  \includegraphics[width=0.98\columnwidth]{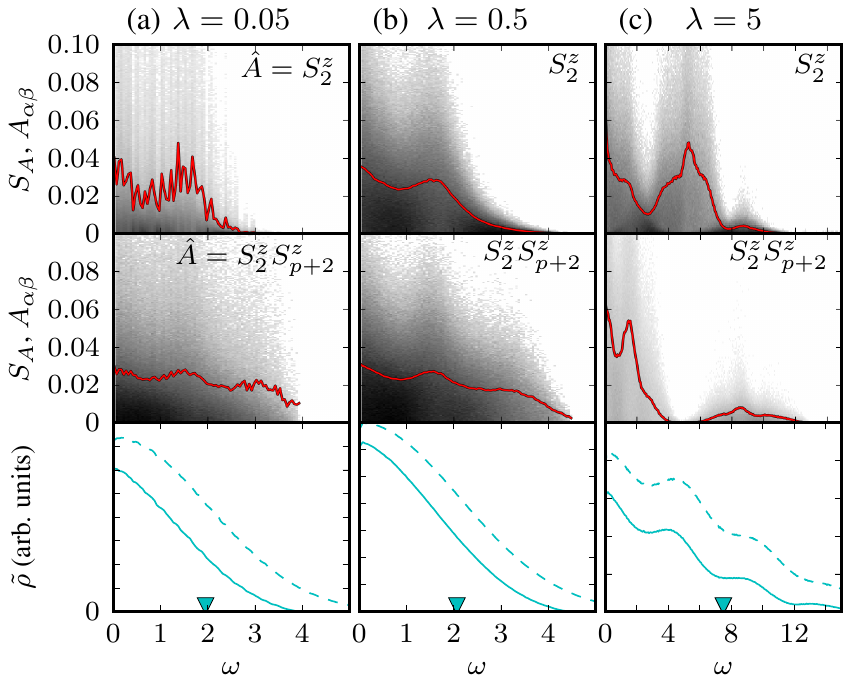}
\caption{  \label{fig_freq}
(Color online) Frequency-resolved analysis of off-diagonal elements for the $(L,\nup)=(13,6)$ ladder.
The shadings show the distribution of the values of $\abs{A_{\alpha\beta}}$ for each frequency
window; darker shading indicates more occurrences of respective $\abs{A_{\alpha\beta}}$ values.
The curves show $S_A$ as a function of the frequency $\omega$. In the bottom panel, we indicate
the density of pairs $\tilde{\rho}$ for the central part of the spectrum (solid) and for the
full spectrum (dashed), in arbitrary units. The marker on the horizontal axis points to the
typical frequency $\omega_0$. The shading and solid curves are all results using the central
part of the spectrum.  The units of frequency are the same as the units used for energy in
Fig.~\ref{fig_Aalphabeta}.  
}
\end{figure}

In the large-system limit, the number of states $\tilde{N}_{\omega,\Delta \omega}$ in the window can be approximated as $\tilde{N}_{\omega,\Delta \omega}\approx2\Delta \omega\tilde{\rho}(E)$, where $\tilde\rho(\omega)$ is the \emph{density of pairs}, i.e., the density of values $E_\alpha-E_\beta$. The density of pairs is defined as the autocorrelation integral
\begin{equation}\label{eqn_tilde_rho}
  \tilde{\rho}(\omega) = \int \rho(E)\rho(E-\omega) d E
\end{equation}
of the density of states $\rho(E)$ with itself.
We note that the density of pairs does not show signatures of the level-spacing statistics, because the density of states is considered on a coarser resolution than that of individual eigenvalues.
The behavior of $\tilde{\rho}(\omega)$ is shown in the bottom row of Figure \ref{fig_freq} for the
spin ladder system, for different values of the $\lambda$ parameter.

To distinguish frequency regimes, we define a ``typical'' frequency scale $\omega_0$, as the
root-mean-square of all possible frequencies:
\begin{equation}\label{eqn_omega0}
  \omega_0^2
  = \var(E_\alpha-E_\beta)
  = \frac{1}{D^2}\sum_{\alpha,\beta}(E_\alpha-E_\beta)^2
  = 2\var(E_\alpha).
\end{equation}
In Fig.~\ref{fig_freq}, the values $\omega_0$ are indicated by  markers on the horizontal (frequency) axes.

%%%%%%%%%%%%%%% FIGURE %%%%%%%%%%%%%%%
%%%%%%%%%%%%%%%%%%%%%%%%%%%%%%%%%%%%%%

The frequency dependence of $S_A(\omega)$ is shown in Fig.~\ref{fig_freq} for the observables
$\hat{A}=S^z_2$ and $\hat{A}=S^z_2S^z_{p+2}$ in the XXZ ladder model.  
In addition, through the shading in the top two panels, we indicate the distribution of the values
of $\abs{A_{\alpha\beta}}$ in each frequency window.
The value of $\Delta \omega=0.05$ used in this figure is a compromise between being sufficiently
small to resolve the details, and having sufficiently many state pairs for good statistics. In the
cases of Fig.~\ref{fig_freq}, the number of state pairs in the window
$[\omega-\Delta\omega,\omega+\Delta\omega]$ is $\sim 10^4$ for $\omega\lesssim \omega_0$.

At high frequencies ($\omega\gg\omega_0$), $S_A(\omega)$ decreases as a function of
$\omega$.  The decrease is rapid; we have found this to be generally exponential or super-exponential
($\sim$ gaussian) with $\omega$; the details vary with the observable and appear to be non-universal.

At medium frequencies, $S_A(\omega)$ typically shows several peaks.  The oscillatory behavior is
more pronounced near integrability, i.e., for small and large $\lambda$.  
We observe typical small-$\lambda$ behavior in Fig.~\ref{fig_freq}(a): The quantity $S_A$ shows
short-scale oscillations, while the density of pairs $\tilde\rho(\omega)$ is smooth.  We conjecture
that the oscillatory behavior in near-integrable systems is due to the Hilbert space being decomposable into many subspaces weakly coupled by the Hamiltonian.  Whenever $\alpha$ and $\beta$ are in different
subspaces, $A_{\alpha\beta}\approx 0$.

At large $\lambda$, the system splits into weakly coupled subspaces which are in addition separated in
energy, as evidenced by the block-like structure in Fig.~\ref{fig_Aalphabeta}(c).  Thus, the peaks
of $S_A$ are accompanied by those in the density of pairs $\tilde\rho(\omega)$. The blocks are
separated by energy $\sim\lambda$, which can be understood from treating the system as uncoupled
dimers in the $\lambda\to\infty$ limit.  These are also the approximate frequencies at which peaks
can be seen in Fig.~\ref{fig_freq}(c).

\section{Distribution of off-diagonal matrix elements}%
\label{sect_twogaussian}

Having described the variance $S_A^2(\omega)$ of the distribution of the values of
$A_{\alpha\beta}$, we now look at the full distribution.

%%%%%%%%%%%%%%% FIGURE %%%%%%%%%%%%%%%
\begin{figure}[t]
\centering
  \includegraphics[width=0.98\columnwidth]{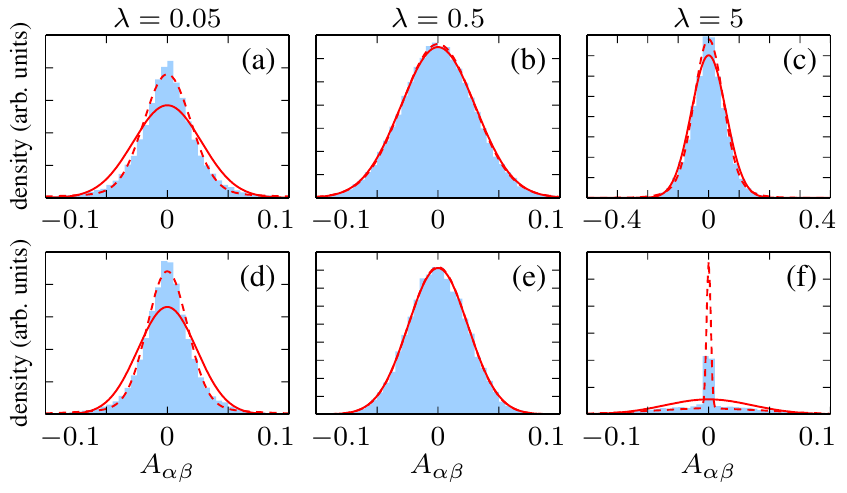}
\caption{   \label{fig_histograms}
(Color online) Histograms (shaded area) of the off-diagonal elements $A_{\alpha\beta}$ with (a--c)
$E_\beta-E_\alpha\in(0,0.05\omega_0)$ and (d--f)
$E_\beta-E_\alpha\in[0.25\omega_0,0.25\omega_0+\delta\omega)$, where $\delta\omega=0.05$. The solid
  curve is a gaussian fit and the dashed curve is a fit of a mixture of two gaussians,
  Eq.\ \eqref{eqn_double_gaussian}.  The observable  is
  $\hat{A}=S^z_2S^z_{p+2}$ and the system size is $(L,\nup)=(13,6)$. The number of state pairs in these
  histogram computations ranges from $4000$ to $50000$.
}
\end{figure}
%%%%%%%%%%%%%%%%%%%%%%%%%%%%%%%%%%%%%%

\subsection{Shapes of the  distributions of  $A_{\alpha\beta}$}

In Fig.~\ref{fig_freq}, we have shown using shading densities the frequency-resolved distributions of
values of $|A_{\alpha\beta}|^2$.  A feature visible already in the density plots is that the
distributions are more strongly weighted near zero (near the horizontal axis) near integrability.
This feature will be explored and described in more detail below.

In Fig.~\ref{fig_histograms}, we show the distributions of $A_{\alpha\beta}$ values, in two
different frequency windows.  The top panels show the low-frequency regime (cutoff frequency
$\omega_\mathrm{max}=0.05\omega_0$).  The bottom panels focus on a frequency window around
$0.25\omega_0$.  Only the states in the central part of the spectrum (within the dashed square
region in Fig.~\ref{fig_Aalphabeta}) are considered.

The distributions are seen to be very nearly symmetric around zero.  Of course, the signs of
individual $A_{\alpha\beta}$ values are not meaningful since every eigenstate carries an arbitrary
phase.  However, from $N\gg1$ eigenstates, one obtains $\frac{1}{2}N^2\gg{N}$ matrix elements; so
the overall shape of the distribution (roughly equal number of positive and negative values) cannot
be altered by the choice of phases for the eigenstates.  

The solid curves are gaussian fits determined by the variance of the $A_{\alpha\beta}$, centered at
$0$.  Far from integrability, this is seen to be a very good description.  However, near
integrability the distribution has a sharper peak than a gaussian, and appears to be a mixture of
two near-gaussian distributions with different widths.  This appears to be a fundamental distinction
between (near-)integrable and generic systems.  

We do not currently have a complete explanation for the extra peak in near-integrable systems, but
we  conjecture the following mechanism which provides some intuition. 
In the integrable case, there are many conserved quantities. The energy eigenstates can be grouped
into subspaces or symmetry sectors by the eigenvalues (``quantum numbers'') of the operators
corresponding to these conserved quantities.  An approximate version of this statement is true close
to, but not at, integrability.
An operator $\hat{A}$ corresponding to a local observable, when acting on an eigenstate
$\ket{\psi_\alpha}$, changes the eigenstate only locally, i.e., slightly.  The resulting wavevector
$\hat{A}\ket{\psi_\alpha}$ will thus be likely to have larger overlap with eigenstates having the
same quantum numbers as $\ket{\psi_\alpha}$, and much smaller overlaps with eigenstates having
different quantum numbers from those of $\ket{\psi_\alpha}$.  In other words, $A_{\alpha\beta}$ is
close to zero whenever $\alpha$ and $\beta$ belong to different subspaces.  Of course,
$\ket{\psi_\alpha}$ and $\ket{\psi_\beta}$ are orthogonal even if they belong to the same subspace,
so that the off-diagonal matrix element of a local operator is small anyway for large system sizes.
The argument is that, when they belong to different sectors, the states differ additionally by
having different quantum numbers, not only by being orthogonal, and this should make the
inter-subspace matrix elements statistically much smaller than intra-subspace matrix elements.

This line of reasoning intuitively connects to the idea that integrablility makes a system
``non-ergodic''.  However, the argument is difficult to make rigorous.  It is easy to construct
special operators that connect different subspaces, e.g., if $\sum_jS_j^z$ is a conserved quantity
in a spin Hamiltonian, the local operator $S_j^+$ will connect different subspaces.  However, a
generic operator is expected not to have such special relationships with many of the conserved
quantities, since most conserved quantities in integrable lattice models have rather complicated
form when expressed in terms of spatially local operators.  Although the explanation provided by
this ``inter-subspace versus intra-subspace'' perspective remains only heuristic at this stage, our
data for multiple system demonstrates that near-integrable systems indeed have a substantial number 
of extremely small matrix elements.

In summary, numerical observations on the families of systems (XXZ ladder, Bose-Hubbard chain)
investigated in this work indicate that, in quasi-integrable cases, the studied local observables
tend to respect the symmetries that are dynamically conserved at exact integrability. We expect this
behavior to be generic at (near-) integrability for local observables in these types of models.
Moreover, we conjecture that this behavior may also be generic in the class of integrable many-body
systems at large, and for a wide class of local observables.

The gaussian shape of the distributions for generic non-integrable points can be explained
heuristically by invoking the central limit theorem. Writing
$c^{(\alpha)}_\gamma\equiv\braket{\phi_\gamma}{\psi_\alpha}$ in terms of the eigenstates
$\ket{\psi_\alpha}$ of the Hamiltonian and $\ket{\phi_\gamma}$ of $A$ (with eigenvalues $a_\gamma$),
we can write the matrix elements as
\begin{equation}\label{eqn_aab_central_limit_theorem}
  A_{\alpha\beta} = \sum_\gamma c^{(\alpha)*}_\gamma c^{(\beta)}_\gamma a_\gamma.
\end{equation}
For non-integrable systems, the summands $c^{(\alpha)*}_\gamma c^{(\beta)}_\gamma a_\gamma$ may be
expected to behave like quasi-independent random variables.  The central limit theorem then implies
the gaussian distribution of $A_{\alpha\beta}$.  As in Ref.\ \cite{BeugelingEA2014PRE}, we stress
that the randomness and independence of coefficients is a hypothesis and difficult to prove
rigorously.  This is in the same spirit as arguments for scaling behaviors of diagonal matrix
elements or of inverse participation ratios based on similar randomness assumptions
\cite{BeugelingEA2014PRE, NeuenhahnMarquardt2012}.
The physical intuition for such randomness assumptions is that an eigenstate in the middle of the
spectrum of a generic system is so complex that the coefficients behave as random and independent
variables for many purposes.

\subsection{Quantifying the  distribution shapes}

In order to characterize the nature of the distributions at small and large $\lambda$, we fit the
numerically obtained histograms to the sum of two gaussian distributions, defined as
\begin{equation}\label{eqn_double_gaussian}
  g(A) = a n_{\sigma_1}(A)+(1-a)n_{\sigma_2}(A),
\end{equation}
where $n_{\sigma_i}(A)$ is the gaussian distribution with variance $\sigma_i^2$ and zero mean, and
$0\leq a \leq 1$.  There are three fit parameters, $a$, $\sigma_1$, and $\sigma_2$ (with $\sigma_1 <
\sigma_2$). Two parameters are determined by equating the variance
$\sigma^2=a\sigma_1^2+(1-a)\sigma_2^2$ and excess kurtosis
$k=\kappa-3=3a(1-a)(\sigma_1^2-\sigma_2^2)^2/\sigma^4$ of $g(A)$ to that of the data. We then
perform a least-squares fit of the cumulative density function of the data to solve for the
remaining degree of freedom $a$. (See Appendix~\ref{app_twogaussian} for details.)

The resulting distributions $g(A)$ are plotted in Fig.~\ref{fig_histograms} as dashed curves.  The
two-gaussian form works very well for small $\lambda$, and reasonably well for large $\lambda$.  The
discrepancy in Fig.~\ref{fig_histograms}(f) may be simply due to the lack of sufficient data points
to provide good statistics for these particular parameters.

%%%%%%%%%%%%%%% FIGURE %%%%%%%%%%%%%%%
\begin{figure}[t]
\noindent\center
\includegraphics[width=0.98\columnwidth]{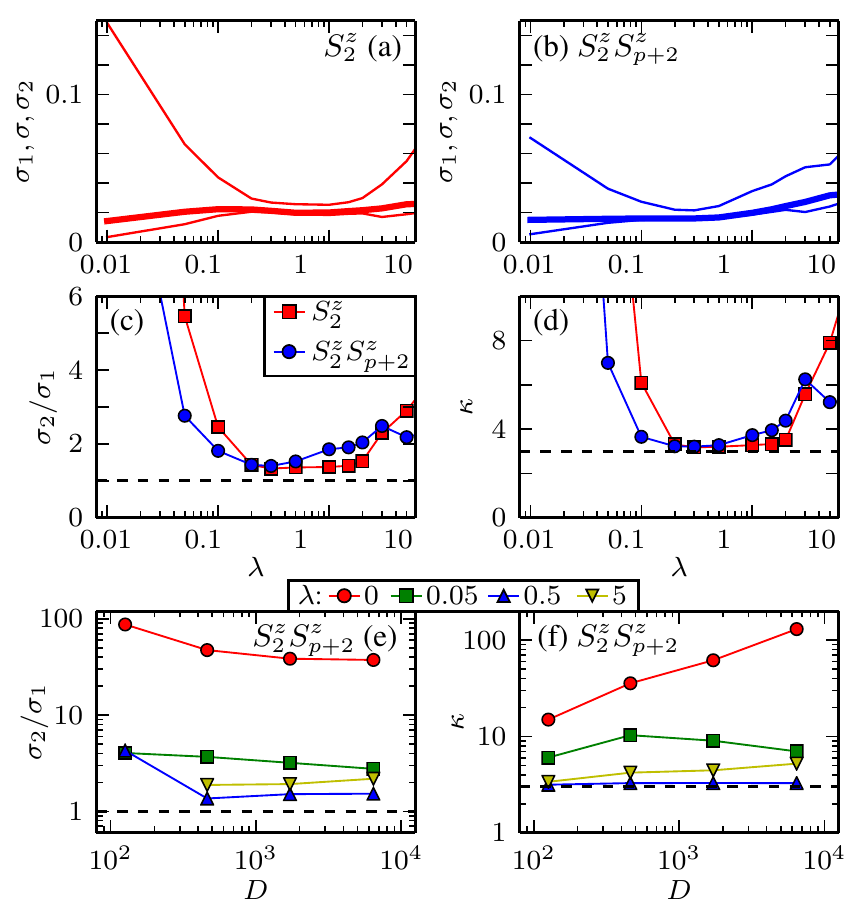}
\caption{
  \label{fig_twogaussian} 
(Color online) Characteristics of the distribution of $A_{\alpha\beta}$.  (a,b) Standard deviations $\sigma_1$,
$\sigma$, and $\sigma_2$ of the ``inner'' gaussian of \eqn\eqref{eqn_double_gaussian}, the full
distribution, and the ``outer'' gaussian, in increasing order.  The thicker curve is
$\sigma$. System size is $(L,\nup)=(15,7)$. (c,d) Ratio $\sigma_1/\sigma_2$ and kurtosis $\kappa$.
The dashed horizontal lines are the values for the gaussian distribution ($\sigma_2/\sigma_1=1$, $\kappa=3$).  We
show results for $\hat{A}=S^z_2$ in red (squares) and for $\hat{A}=S^z_2S^z_{p+2}$ in blue
(circles).  (e,f) $\sigma_2/\sigma_1$ and $\kappa$ as a function of the Hilbert-space dimension $D$
for several values of $\lambda$.  In (e), data for the smallest system size for $\lambda=5$ is
absent --- the  procedure does not yield a solution for $\sigma_i$ due to the low
density of states. 
}
\end{figure}
%%%%%%%%%%%%%%%%%%%%%%%%%%%%%%%%%%%%%%

In Fig.~\ref{fig_twogaussian}, we show data related to this two-component description ($\sigma_{1,2}$,
$\sigma$, $\kappa$), for the observables $S^z_2$ and $S^z_2S^z_{p+2}$ in the ladder system.  The two
standard deviations generally become equal at intermediate $\lambda$ (the ratio $\sigma_2/\sigma_1$
drops to near unity), indicating that a single-gaussian description works well away from
integrability.
In (d), we show the kurtosis $\kappa$ of the distribution, used as an input for the fit.  The
kurtosis is close to $3$ (the kurtosis value of the gaussian distribution) in the intermediate regime,
again showing that a single gaussian is a good description for the distribution of $A_{\alpha\beta}$
values in generic systems.
The kurtosis is significantly larger than 3 as one approaches the integrable points, signifying a
stronger central peak than that of a single gaussian.

In Figs.~\ref{fig_twogaussian}(e) and (f), we provide a scaling analysis by plotting $\sigma_2/\sigma_1$
and $\kappa$ as a function of the Hilbert-space dimension $D$ for the observable
$\hat{A}=S^z_2S^z_{p+2}$.  In the non-integrable regime, the values remain near
$\sigma_2/\sigma_1\approx1$ and $\kappa\approx3$ as the sizes are increased.  For $\lambda=0$, the
kurtosis $\kappa$ increases away from 3 with larger $D$, indicating that the central peak gets
stronger relative to the larger gaussian as the system size increases.  This is consistent with our
explanation of the two-component structure in terms of symmetry sectors: the number of eigenstate
pairs belonging to different subspace increases faster with $D$ compared to the number of eigenstate
pairs within the same symmetry subspace.

Also noteworthy is the behavior at the near-integrable point $\lambda=0.05$: the data shows
convergence with increasing $D$ toward the non-integrable values $\sigma_2/\sigma_1=1$ and
$\kappa=3$.  In particular $\kappa$ shows non-monotonic behavior: it first increases like in the
integrable case, and only beyond a certain size starts decreasing back toward the single-gaussian
value $\kappa=3$.  This is a manifestation of the phenomenon that, near but not exactly at
integrability, the system size needs to be large to show generic non-integrable behavior
\cite{BeugelingEA2014PRE}.

\section{Scaling analysis}
\label{sect_scaling}%

In this section we analyze the system-size dependence of the average magnitudes of
$\abs{A_{\alpha\beta}}^2$, which corresponds to the widths of the distributions studied in the
previous section. 

The average value of $\abs{A_{\alpha\beta}}^2$ close to the diagonal in the central part of the
spectrum (omitting the lowest and highest $25\%$ of the energy range, as indicated by the dashed
squares in Fig.~\ref{fig_Aalphabeta}) is given by
\begin{equation}\label{eqn_gammabar}
\gammabar ~=~
  \frac{1}{\tilde{\mathcal{N}}}\sum^{\sim}_{\substack{\alpha,\beta\\\alpha\not=\beta\\\abs{E_\beta-E_\alpha}\leq\omega_\mathrm{max}}}
  \abs{A_{\alpha\beta}}^2  ~=~  S_A^2(0,\omega_{\mathrm{max}}) .
\end{equation}
Here, $\sum\limits^{\sim}$ denotes summation over the relevant state pairs: It includes all $\alpha$ and
$\beta$ within the bulk of the spectrum with $\alpha\not=\beta$ and with $\abs{E_\alpha-E_\beta}\leq
\omega_\mathrm{max}$, where $\omega_\mathrm{max}$ acts as the frequency cutoff. 

The quantity $\gammabar$ depends on the cutoff frequency $\omega_\mathrm{max}$.  We consider two
values of $\omega_\mathrm{max}$. First, we define a low-frequency measure, $\gammabar_\mathrm{low}=
\gammabar(\omega_\mathrm{max}= 0.05\omega_0)$, where $\omega_0$ is the ``typical frequency''
[\eqn\eqref{eqn_omega0}].  Second, we define $\gammabar_\mathrm{all}=
\gammabar(\omega_\mathrm{max}\to\infty)$ including all state pairs within the bulk of the spectrum
(dashed square in Figure \ref{fig_Aalphabeta}).

%%%%%%%%%%%%%%% FIGURE %%%%%%%%%%%%%%%
\begin{figure}[t]
  \noindent\center%
  \includegraphics[width=0.98\columnwidth]{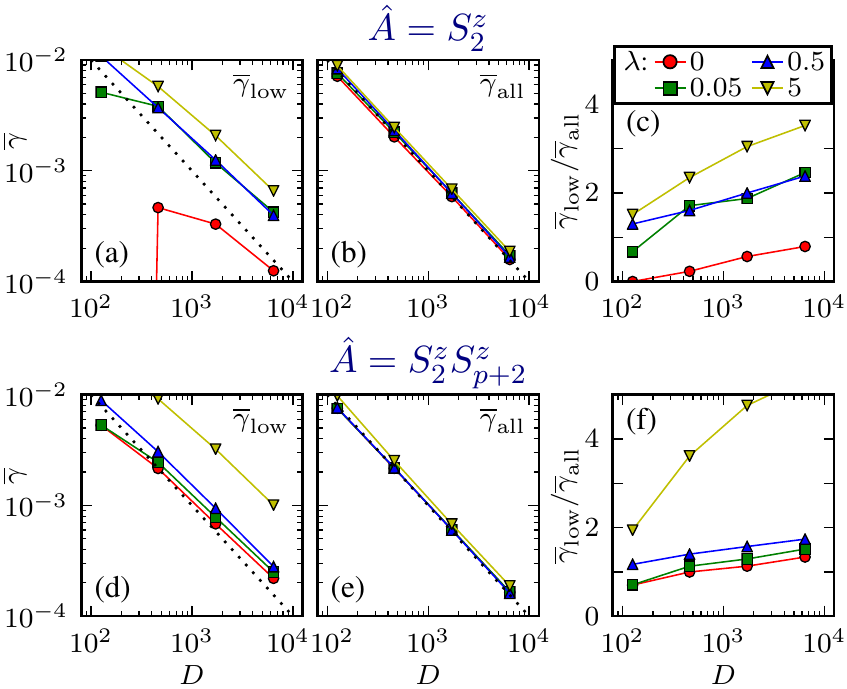}
  \caption{  \label{fig_lowfreq_systemsize}
(Color online) System-size scaling analysis of average $\abs{A_{\alpha\beta}}^2$ through the quantities (a)
$\gammabar_\mathrm{low}$, (b) $\gammabar_\mathrm{all}$, and (c)
$\gammabar_\mathrm{low}/\gammabar_\mathrm{all}$ for the observable $\hat{A}=S^z_2$, for several
values of $\lambda$.  The respective results for $\hat{A}=S^z_2S^z_{p+2}$ are shown in panels
(d--f). The dotted lines in (a), (b), (d) and (e) are $\gammabar=1/D$.
}
\end{figure}
%%%%%%%%%%%%%%%%%%%%%%%%%%%%%%%%%%%%%%

In Fig.~\ref{fig_lowfreq_systemsize}, we plot the quantities $\gammabar_\mathrm{low}$,
$\gammabar_\mathrm{all}$, and the ratio $\gammabar_\mathrm{low}/\gammabar_\mathrm{all}$ as a
function of Hilbert-space size $D$ for several values of $\lambda$, in the top row for the
observable $\hat{A}=S^z_2$ and in the lower row for $\hat{A}=S^z_2S^z_{p+2}$.  Both
$\gammabar_\mathrm{low}$ and $\gammabar_\mathrm{all}$ show a power-law behavior, $\propto D^{-1}$.
the scaling is almost exact for $\gammabar_\mathrm{all}$.  This scaling behavior is consistent with
the scaling of the temporal fluctuations being exponential in $L$, as observed in
Ref.~\cite{ZangaraEA2013}.  

The $D^{-1}$ scaling for non-integrable systems can be explained by using the central limit theorem
invoked in the previous section to explain the gaussian form of the distribution of
$A_{\alpha\beta}$ values.  From \eqn\eqref{eqn_aab_central_limit_theorem}, we interpret
$A_{\alpha\beta}$ as the average of the random variables $X_\gamma\equiv Dc^{(\alpha)*}_\gamma
c^{(\beta)}_\gamma a_\gamma$.  Assuming $c^{(\alpha)}_\gamma$ and $c^{(\beta)}_\gamma$ to be
independent random variables, each with variance $1/D$ due to normalization of the eigenfunctions,
the random variables $X_\gamma$ can be argued to be independent and to have $D$-independent
variance, $\var(X_\gamma)\sim 1$.  The central limit theorem then states that the variance of
$A_{\alpha\beta}$ (i.e., the average of $\abs{A_{\alpha\beta}}^2$) scales as
$\var(X_\gamma)/D\sim1/D$.  As in Ref.~\cite{BeugelingEA2014PRE}, the argument relies or
difficult-to-prove randomness assumptions.

The $D^{-1}$ scaling can be more directly understood by estimating the average value of all
$\abs{A_{\alpha\beta}}^2$, including the edges of the spectrum and the diagonal elements, which is
equal to $\Tr(A^2)/D^2$.  For local observables, $\Tr(A^2)\propto D$.  (In fact, for the two
observables in Fig.~\ref{fig_lowfreq_systemsize}, $\Tr(A^2)= D$ exactly.)  The scaling of the
average as $\propto D^{-1}$ immediately follows.  Of course, both $\gammabar_\mathrm{all}$ and
$\gammabar_\mathrm{low}$ are slightly different from $\Tr(A^2)/D^2$. For $\gammabar_\mathrm{all}$,
the states outside the central part and the diagonal elements are not included, as opposed to
$\Tr(A^2)/D^2$ where they are included.  Nevertheless, in Fig.~\ref{fig_lowfreq_systemsize}(b,e),
$\gammabar_\mathrm{all}$ (data points) follows $\Tr(A^2)/D^2=1/D$ (dotted line) very closely, for
all $\lambda$.  This shows that the contribution from the diagonal elements and from the edge states
are negligible.

In Fig.~\ref{fig_lowfreq_systemsize}(a,d), $\gammabar_\mathrm{low}$ shows approximate $\propto
D^{-1}$ scaling. The magnitudes are generally larger than $1/D$ for larger $D$, reflecting the fact
that the low-frequency $\abs{A_{\alpha\beta}}$ are on average larger than other off-diagonal matrix
elements (as seen previously in Figs.\ \ref{fig_Aalphabeta} and \ref{fig_freq}).  This is also
reflected, Fig.~\ref{fig_lowfreq_systemsize}(c,f), in the ratio
$\gammabar_\mathrm{low}/\gammabar_\mathrm{all}$.  The ratio $>1$ for larger sizes.  The effect is
most prominent for large $\lambda$, which reflects the very large concentration near the diagonal
seen in Figs.\ \ref{fig_Aalphabeta}(c) and \ref{fig_freq}(c).
The ratios $\gammabar_\mathrm{low}/\gammabar_\mathrm{all}$ increase with system size.  It is
conceivable that these ratios will converge to a constant at larger $D$, so that
$\gammabar_\mathrm{low}$ also converges to a $\propto D^{-1}$ dependence. The available data hints
at such behavior, but the available systems sizes are insufficient to make a definitive statement.

The scaling of $\gammabar_\mathrm{low}$ deviates from the $\propto D^{-1}$ especially for smaller
systems and close to integrability. This behavior is reminiscent of the fluctuations of
$A_{\alpha\alpha}$ close to (but not exactly at) an integrable point, where the scaling deviates
from $D^{-1/2}$ for intermediate sizes but converges to $D^{-1/2}$ as the system size is increased
\cite{BeugelingEA2014PRE}.

\section{Bose-Hubbard chain}
\label{sect_bosehubbard}

To evaluate the generality of the results presented in previous sections with the XXZ ladder system,
we present in this section a summary of analogous data for the Bose-Hubbard chain,
\eqn\eqref{eqn_h_bosehubbard}.  We show data for the observable $\hat{A}=b_2^\dagger b_3 +
b_3^\dagger b_2$.

%%%%%%%%%%%%%% FIGURE %%%%%%%%%%%%%%%
\begin{figure}[t]
  \includegraphics[width=0.98\columnwidth]{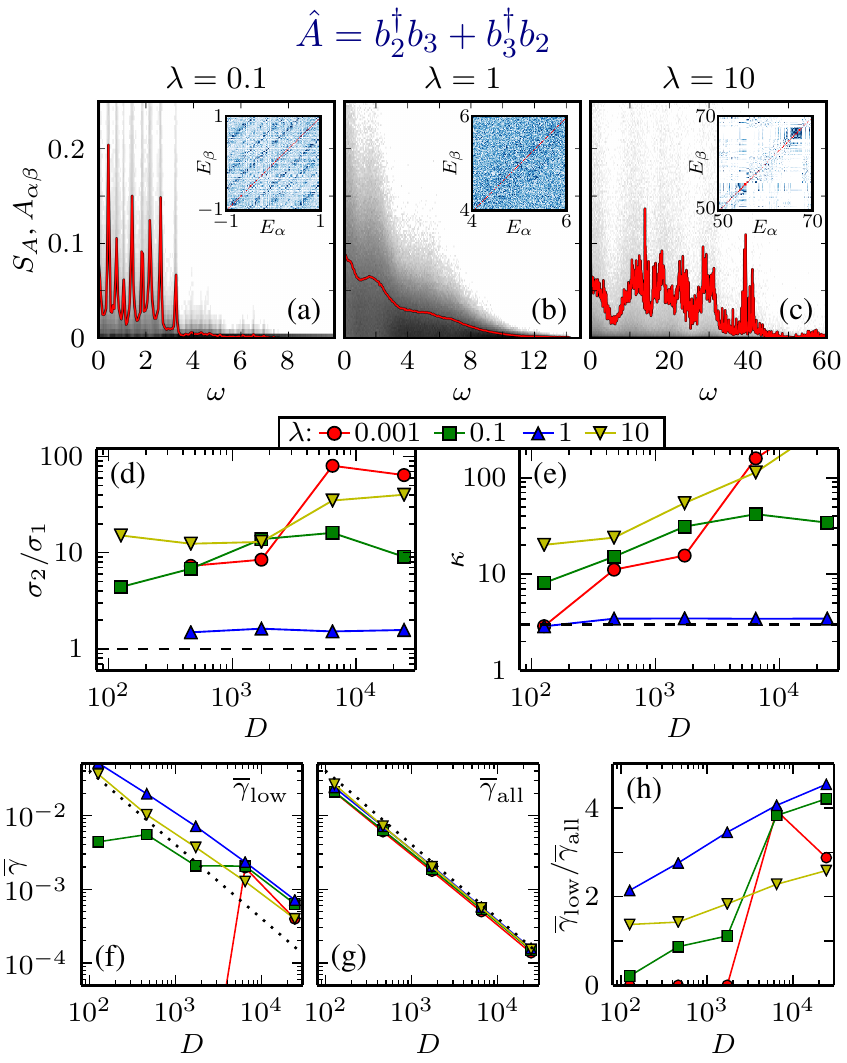}
  \caption{   \label{fig_bh}
(Color online) Bose-Hubbard model; observable $\hat{A}=b_2^\dagger b_3+b_3^\dagger b_2$.  (a,b,c)
Frequency-dependence shown through $S_A$, as in Fig.~\ref{fig_freq}.  Insets show fragments of
the density plot of $\abs{A_{\alpha\beta}}$ as function of $E_\alpha$ and $E_\beta$, as in
Fig.~\ref{fig_Aalphabeta}. The system size is $(L,\nb)=(7,7)$.  (d,e) Analysis of the
shape of the distribution through $\sigma_2/\sigma_1$ and kurtosis $\kappa$, as in Fig.~\ref{fig_twogaussian}.
(f,g,h) Average matrix element $\gammabar_\mathrm{low}$ and $\gammabar_\mathrm{all}$ for low and all
frequencies, and their ratio, as in Fig.~\ref{fig_lowfreq_systemsize}. In (f,g), the dotted
lines are  $\gammabar=4/D$.
}

\end{figure}
%%%%%%%%%%%%%%%%%%%%%%%%%%%%%%%%%%%%%

The frequency-resolved analysis of the matrix elements $A_{\alpha\beta}$ is performed for the values
$\lambda=1$, typical for the nonintegrable regime, and $\lambda=0.1$ and $10$ close to the two
integrable limits.  The results, in Figs.~\ref{fig_bh}(a)--(c), are qualitatively similar to the
ones for the XXZ model in Fig.~\ref{fig_freq}.   

In Figs.~\ref{fig_bh}(d) and (e), we analyze the distribution of the values $A_{\alpha\beta}$ for
low frequencies, by fitting to the two-component distribution as described in Sect.~\ref{sect_twogaussian}.
The ratio $\sigma_2/\sigma_1$ and the kurtosis $\kappa$ are high ($\gg 1$ and $\gg 3$, respectively)
at or near integrability.  In the nonintegrable regime (represented by $\lambda=1$), both quantities
are close to the values appropriate for a gaussian distribution ($1$ and $3$, respectively).

In Figs.~\ref{fig_bh}(d)--(h) we show data for $\lambda=10^{-3}$ as a substitute for the exact
integrable point $\lambda=0$, because the strong oscillations at $\lambda=0$ make our procedure for
extracting $\sigma_{1,2}$ (Appendix \ref{app_twogaussian}) unreliable.  At accessible sizes, the
$\lambda=10^{-3}$ data indeed shows size-dependence characteristic of integrable points: increase of
$\kappa>3$ with increasing system size.  At some very large system size, $\kappa(D)$ is expected to
decrease again.  Such nonmonotonic behavior is a signature of proximity to integrability.  The
non-monotonic behavior is visible at available system sizes for the $\lambda=0.1$ data.

The root-mean-squared $\gammabar_\mathrm{all}$ [Fig.~\ref{fig_bh}(g)] of the matrix elements
$\abs{A_{\alpha\beta}}^2$ without frequency cutoff shows a $D^{-1}$ scaling, the values being close
to $\Tr(A^2)/D^2=4/D$
\footnote{Here, $\Tr(A^2)=\binom{L+\nb}{L-1}=2D(L+\nb)\nb/L(L+1)$ which equals $4DL/(L+1)$ if
  $\nb=L$, i.e., at unit filling.}.
The scaling of $\gammabar_\mathrm{low}$ is not equally clear at these sizes.
The erratic behavior close to integrability is possibly due to the presence of many very sharp peaks
in $S_A$, especially at low frequencies [see Fig.~\ref{fig_bh}(a)].

\section{Discussion}
\label{sect_discussion}

Motivated by the importance of off-diagonal matrix elements ($A_{\alpha\beta}$) of local operators
in the physics of time evolution after a quantum quench, we have provided a detailed study of the
statistical properties of such matrix elements, for systems with short-range interactions.  Data on
off-diagonal matrix elements have appeared in the non-equilibrium literature (e.g.\
\cite{SteinigewegEA2013, KhatamiEA2013, RigolEA2008}); the present work extends such work to provide
a systematic account of these objects.
We have chosen multiple observables and families of Hamiltonians, and have thus been able to extract general features.  We have also elucidated the role of proximity to integrability as well as the approach to the thermodynamic limit. 

The distribution of values of $A_{\alpha\beta}$ is gaussian for generic systems, but deviates in a
particular way (stronger peak at zero, or mixture of two gaussian-like distributions) as one
approaches integrability.  We have used this to formulate a quantitative characterization of
proximity to integrability, through the kurtosis $\kappa$ of the distribution.  We find
$\kappa\sim3$ for non-integrable (generic or chaotic) systems, and a larger $\kappa$ that
\emph{increases with system size} for integrable systems.  This distinction makes it possible to
graphically represent our idea, formulated in Ref.~\cite{BeugelingEA2014PRE}, that distance from
integrability can be characterized by a length scale ---for near-integrable systems, the
size-dependence $\kappa(D)$ shows an initial increase followed by a decrease beyond a certain size.
This size $D$ where $\kappa(D)$ is maximal characterizes the proximity to integrability, and
increases as one approaches integrability. These properties may prove to be useful signatures for
proximity to integrability, in the sense that they can be determined straightforwardly from Fourier
transforms of the time evolution after a quench, which may be feasible to observe in experiments.

The average magnitude of the matrix elements, $S^2(\omega)$ or $\gammabar$, determines the magnitude
of temporal fluctuations of $\avg{A}(t)$ after a quantum quench.  The scaling of this quantity for
non-integrable systems, $\sim{1/D}$, is consistent with the scaling of temporal fluctuations known
from the literature \cite{ZangaraEA2013}.  We also find that the low-frequency average is higher
than the average over all frequencies, $\gammabar_\mathrm{low}>\gammabar_\mathrm{all}$ (Figs.\
\ref{fig_lowfreq_systemsize},\ref{fig_bh}), reflecting the overall decrease of $S(\omega)$ with
increasing frequency (Fig.~\ref{fig_freq}).  This suggests that, for quenches in nonintegrable
systems, low-frequency contributions are likely to dominate in the time evolution, regardless of
whether or not the initial conditions are very local in energy.

The $\sim{1/D}$ scaling can be argued from the central limit theorem assuming wavefunction
coefficients of non-integrable Hamiltonians to be effectively random.  This is a recurring
assumption in this field (e.g., \cite{BeugelingEA2014PRE, ZangaraEA2013, NeuenhahnMarquardt2012}),
usually without rigorous proof, but with a similar status as the ETH, namely, as a plausible hypothesis, the validity of which has to be established by numerical results. Nevertheless, this argument is useful because it also provides a plausible explanation for the gaussian distributions of the matrix elements. For the scaling, we have provided an alternate argument based on the trace
of local operators, which turns out to work well especially for $\gammabar_\mathrm{all}$.

The present work raises a number of questions for further study.  As a new characterization of
integrability, the double-peak structure of the $A_{\alpha\beta}$ distribution deserves to be better
understood.  The relative weight of the inner peak is presumably connected to the distribution of
sizes of the many subspaces that the Hilbert space is divided into, due to the many conservation laws
present at integrability.  At present, we do not have a quantitative understanding of the exact
connection between the subspace distribution and the non-gaussian distribution of the off-diagonal
matrix elements, although the presence of many subspaces provides a plausible explanation for the
double-peak form.  
A related question is the type of deviation from the gaussian shape of
the $A_{\alpha\beta}$ distribution for systems with a few (nonzero but $O(L^0)$) conservation laws.
It would also be interesting to find out whether the two-component versus gaussian
(single-component) structures can be related to differences in real-time relaxation and fluctuation
behaviors between integrable and non-integrable systems.  Also, it is possible that our findings for
near-integrable points might have consequences for ``pre-thermalization'' behaviors
\cite{MoeckelKehrein2008, EcksteinKollar2008,KollarEA2011,ZhangEA2012PRE,MarcuzziEA2013}.

\acknowledgments

We thank J.~Dubail, P.~Ribeiro, L.~Santos, and J.~M.~St{\'e}phan for interesting discussions.

\appendix
\section{Time evolution and off-diagonal matrix elements}%
\label{app_time_evolution}

In this Appendix we outline some of the connections to time evolution which motivates the study of
off-diagonal matrix elements.  
We consider a isolated quantum system with Hamiltonian $H$, with eigenvalues $E_\alpha$ and
eigenstates $\ket{\psi_\alpha}$.  Under this Hamiltonian, the time evolution of the initial state
$\ket{\Psi(0)}$, that may be the result of a quench at $t=0$ from another Hamiltonian, is given by
$\ket{\Psi(t)}=\sum_\alpha c_\alpha\ee^{-\ii E_\alpha t}\ket{\psi_\alpha}$, 
where $c_\alpha=\braket{\psi_\alpha}{\Psi(0)}$ are the expansion coefficients in the eigenstate
basis.  Given an observable $A$, its expectation value evolves as 
\begin{equation}\label{eqn_time_expval}
  \avg{A}(t)
   = \bramidket{\Psi(t)}{\hat{A}}{\Psi(t)}
   = \sum_{\alpha,\beta}c_\alpha^*c_\beta A_{\alpha\beta}\ee^{\ii (E_\alpha-E_\beta)t} .
\end{equation}
The long-time average of this quantity is
\begin{equation}\label{eqn_lta}
  \avglta{A} = \lim_{T\to\infty}\frac{1}{T}\int_0^T\avg{A}(t) dt.
\end{equation}
For a non-degenerate spectrum, the off-diagonal terms do not contribute, so that 
$\avglta{A}=\sum_\alpha \abs{c_\alpha}^2A_{\alpha\alpha}$.

While the $A_{\alpha\alpha}$ determine the long-time average, these diagonal matrix elements do not
say anything about the temporal fluctuations $f_A(t) \equiv \avg{A}(t)-\avglta{A}$ around the
average. A representative value for the magnitude of temporal fluctuations is its root-mean-square
\begin{align}
  (\sigma^{\mathrm{t}}_A)^2
   \equiv \overline{[f_A(t)]^2}
   = \lim_{T\to\infty}\frac{1}{T}\int_0^T[f_A(t)]^2 dt  . 
  \label{eqn_time_integral}
\end{align}
Using  \eqn\eqref{eqn_time_expval}, one
finds that
\begin{equation}\label{eqn_time_fluctuations}
  (\sigma^{\mathrm{t}}_A)^2
   =\sum_{\substack{\alpha,\beta\\ \alpha\not=\beta}} \abs{c_\alpha}^2 \abs{c_\beta}^2 \abs{A_{\alpha\beta}}^2,
\end{equation}
under the assumption that the spectrum is incommensurate, i.e., when there are no degeneracies and $E_\alpha+E_\beta=E_\gamma+E_\delta$ implies that $(\alpha,\beta)=(\gamma,\delta)$ or $(\alpha,\beta)=(\delta,\gamma)$.

The fluctuation amplitude $(\sigma^{\mathrm{t}}_A)^2$ can be considered as a correlator 
of $f_A(t)$ with itself.  Generalizing to correlators at different times, we get the autocorrelation
function, 
\begin{equation}\label{eqn_autocorrelation_time}
  \overline{f_A(t)f_A(t+\tau)}
 =\sum_{\substack{\alpha,\beta\\ \alpha\not=\beta}} \abs{c_\alpha}^2 \abs{c_\beta}^2 \abs{A_{\alpha\beta}}^2 \ee^{\ii(E_\beta-E_\alpha)\tau},
\end{equation}
which appears in formulations of non-equilibrium fluctuation-dissipation relations
\cite{KhatamiEA2013, Srednicki1999}.  The Fourier transform of this quantity is 
\begin{equation}\label{eqn_autocorrelation_freq}
  s^2(\omega) =
  \sum_{\substack{\alpha,\beta\\ \alpha\not=\beta}} \abs{c_\alpha}^2 \abs{c_\beta}^2
  \abs{A_{\alpha\beta}}^2 \delta\left(\omega-(E_\beta-E_\alpha)\right) .
\end{equation}
The strength of the fluctuations at frequency $E_\beta-E_\alpha$ is equal to
$\abs{c_\alpha}^2\abs{c_\beta}^2\abs{A_{\alpha\beta}}^2$.

Eqs.\ \eqref{eqn_time_fluctuations} and \eqref{eqn_autocorrelation_freq} demonstrate the roles of
$A_{\alpha\beta}$ in real-time considerations.  The quantity $\gammabar$ in our work can be regarded
as a general version of the right hand side of \eqref{eqn_time_fluctuations} which is independent of
any particular quench protocol.  The quantity $S^2(\omega)$ is similarly a smoothed version of the
right hand side of \eqref{eqn_autocorrelation_freq}, again omitting reference to specific initial
states.

\section{Fit to the distribution of $A_{\alpha\beta}$}
\label{app_twogaussian}%

In Sec.~\ref{sect_twogaussian}, we have fitted the sum of two gaussian distributions $g(A)$
[\eqn\eqref{eqn_double_gaussian}] to the actual distribution $d(A)$ of the off-diagonal elements in
a small frequency window. The fit parameters in this distribution are $a$, the mutual weight of the
two terms, and $\sigma_1$ and $\sigma_2$, the standard deviations. For the fits shown in
Fig.~\ref{fig_histograms} and for the data plotted in Fig.~\ref{fig_twogaussian}, we impose that the
fitted distribution $g(A)$ has the same variance $\sigma^2$ and kurtosis $\kappa$ as the actual
data. This yields the equations $\sigma^2=a\sigma_1^2+(1-a)\sigma_2^2$ and $\kappa - 3 =
3a(1-a)(\sigma_1^2-\sigma_2^2)^2/\sigma^4$. By solving these equations for given $\sigma^2$ and
$k=\kappa-3$, we obtain expressions for $\sigma_1$ and $\sigma_2$ in terms of $a$,
\begin{equation}
\sigma_{1,2}^2 ~=~ \sigma^2\left(1\mp\sqrt{\tfrac{1}{3}k(1-a)/a}\right) \, .
\label{eqn_sigma12_solution}
\end{equation}
We have imposed $\sigma_1^2 \leq \sigma^2 \leq \sigma_2^2$. The remaining variable $a$ can be
obtained in several ways. For Figs.~\ref{fig_histograms} and \ref{fig_twogaussian}, we have obtained
$a$ by numerically minimizing the integrated square difference between the cumulative density
function of $g(A)$ and that of the data $d(A)$. This method yields an ``optimal'' value of $a$,
which is substituted into \eqn\eqref{eqn_sigma12_solution} in order to obtain $\sigma_1$ and
$\sigma_2$.  However, when the cumulative density distribution of the data behaves erratically due
to very few states being involved, this procedure might fail and give an optimal value of $a$
outside the range $[0,1]$ (e.g., the $\lambda=5$ data in Fig.\ \ref{fig_twogaussian}).

%\bibliography{../_ed}
%\bibliographystyle{apsrev4-1}

%merlin.mbs apsrev4-1.bst 2010-07-25 4.21a (PWD, AO, DPC) hacked
%Control: key (0)
%Control: author (72) initials jnrlst
%Control: editor formatted (1) identically to author
%Control: production of article title (-1) disabled
%Control: page (0) single
%Control: year (1) truncated
%Control: production of eprint (0) enabled
%

\end{document}